\documentclass[aps,superscriptaddress,amsmath,amssymb,floatfix,twocolumn,showpacs,amsfonts,longbibliography,prl]{revtex4-1}
\usepackage{times}
\usepackage[varg]{txfonts}
\usepackage{textcomp}
\usepackage{graphicx}
\usepackage{subfigure}
\usepackage{tabu}
\usepackage{color}
\usepackage[colorlinks=true,citecolor=blue,urlcolor=blue,linkcolor=blue,hyperindex]{hyperref}
\usepackage{braket}
\usepackage{float}
\usepackage{overpic}
\usepackage{amssymb}
\usepackage{amsmath}
\usepackage{gensymb}
\usepackage{mathrsfs}
\usepackage{bm}

\allowdisplaybreaks

\begin{document}
%%%%%%%%%%%%%%%%%%%%%%%%%%%%%%%%

\title{Dynamics in the planar pyrochlore lattice: bow-tie flat band and mixed 't Hooft anomaly}
\author{Zijian Xiong}
\affiliation{Department of Physics, and Center of Quantum Materials and Devices, Chongqing University, Chongqing, 401331, China}
\affiliation{Chongqing Key Laboratory for Strongly Coupled Physics, Chongqing, 401331, China}
\author{Yining Xu}
\email{Corresponding author: xuyining@cqnu.edu.cn}
\affiliation{College of Physics and Electronic Engineering, Chongqing Normal University, Chongqing 401331, China}
\author{Xue-Feng Zhang}
\affiliation{Department of Physics, and Center of Quantum Materials and Devices, Chongqing University, Chongqing, 401331, China}
\affiliation{Chongqing Key Laboratory for Strongly Coupled Physics, Chongqing, 401331, China}

\begin{abstract}
The quantum phase transition between $\mathbb{Z}_{2}$ plaquette valence bound solid (PVBS) and superfluid (SF) phases on the planar pyrochlore lattice (square ice) is under debate, because the conventional deconfined theory does not support continuous one, but the numerical evidence is still not solid. Here, we propose the system can be effectively described by a 2+1 dimensional Abelian-Higgs model which may host the mixed 't Hooft anomaly, so that the deconfinement can exist on the domain walls at the transition point. To verify it, we study the spin excitation spectra by combining stochastic analytic continuation and quantum Monte Carlo simulation. In both PVBS and SF phases, a flat band with bow-tie structure is observed and can be explained by group theoretic analysis. At the transition point, the spectra turn to be continuous and gapless, which indicates the topological excitations. From the snapshot of the spin configuration in real space, we found the existence of the domain wall in which the symmetries satisfy the anomaly. Meanwhile, a Luttinger liquid like continuum implies additional domain walls (point-defect) can emerge in the domain walls (line-defect) and take the role of deconfinement at transition point, as prediction of mixed 't Hooft anomaly. Our work can build a new bridge between the topological gauge field theory and a strongly correlated system.
\end{abstract}

\maketitle
\date{\today}
%%%%%%%%%%%%%%%%%%%%%%%%%%%%%%%%%
%Geometry frustration and gauge field
\emph{Introduction.}---Gauge theories play a vital role in particle physics, but also become a powerful tool to study condensed matter physics. As a fertile ground for seeking the unconventional phases and exotic excitations \cite{lacroix2011book,RM2006,Vojta_2018}, the geometrical frustration imposes the strongly local constraint for the ground states and leads to the extensive degeneracy, so that the system can be naturally described by the emergent lattice gauge field \cite{Vojta_2018}. As a paradigmatic example, the pyrochlore quantum spin ice manifesting emergent U(1) gauge structure \cite{Hermele2004prb,Gingras2014review}, and the corresponding effective field theory is QED$_4$. In the past decades, many exotic phenomena in spin ice have been discovered, such as magnetic monopole \cite{Moessner2008nature}, gauge photon \cite{Benton2012prb}, Coulomb phase \cite{Henley2010} and the large emergent fine structure constant \cite{RM2021}.

\begin{figure}[t]
	\centering
	{\subfigure[]{
			\includegraphics[scale=0.6]{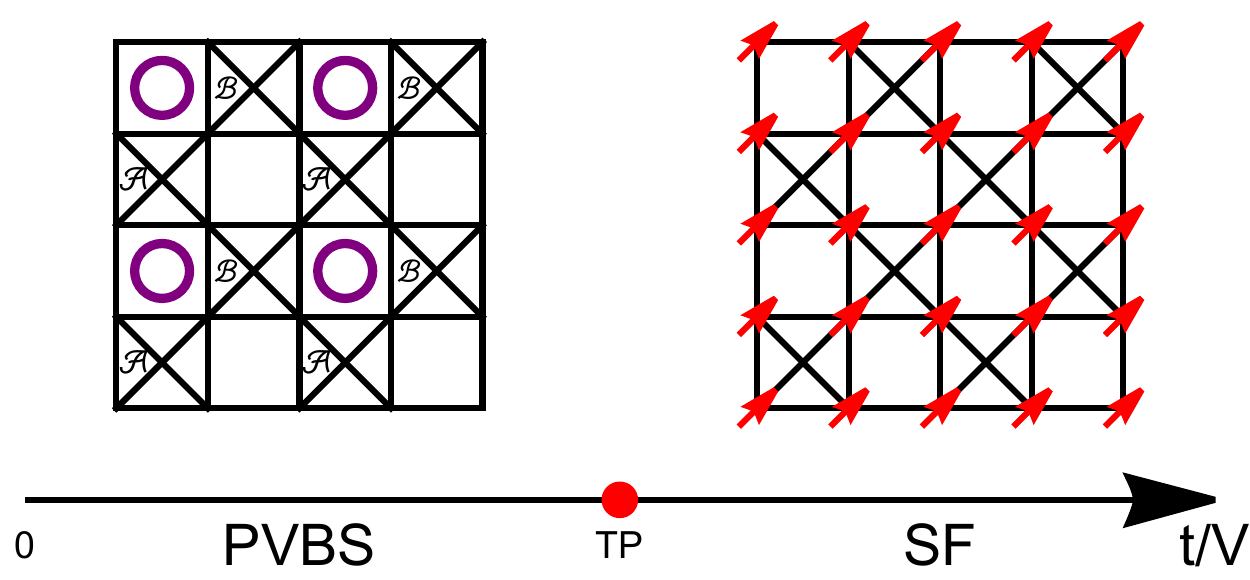}}}
	{\subfigure[]{
			\includegraphics[scale=0.37]{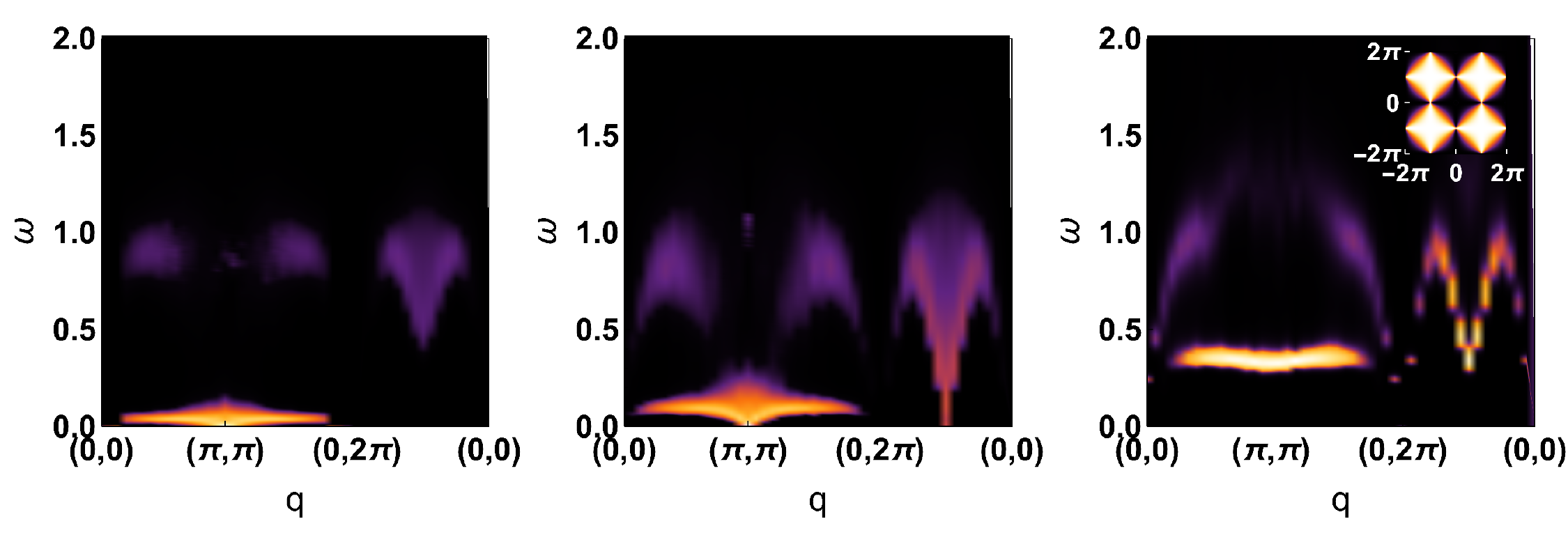}}}
	\caption{\label{fig1} (a) Phase diagram and the schematic depiction of PVBS (left) and SF (right) phases. The flippable plaquettes are denoted by purple circles, and the centers of the crossing plaquette compose the dual lattice with sublattices $\mathcal{A}$ and $\mathcal{B}$. The TP is marked with a red point. (b) The spectra of the PVBS phase (left), TP (middle) and SF (right) phase (Inset: the bow-tie flat band calculated by spin wave theory with constant energy cut).}
\end{figure}

%The planar pyrochlore
In comparison, the effective field theory depicting two dimensional frustrated system could be QED$_3$, and a straightforward example is the two dimensional spin ice -- planar pyrochlore lattice (Fig.\ref{fig1}). The Ising model in this lattice is equivalent to the six-vertex model and exactly solved by Lieb \cite{Lieb1967prl},  the ground state is extensive degenerate and all correlation functions are found to be algebraically decay. When the quantum fluctuation is introduced \cite{Moessner2004,YWan2012prl,Henry2014prl,Ralko2010prl,Shannon2004prb}, the degeneracy will be lifted and the ground state is PVBS which breaks the translational symmetry. On the other hand, in the free particle limit, a flat band physics will naturally emerge due to the lattice geometry \cite{Mielke2015}. However, the phase transition between them is still controversial \cite{Ralko2010prl,Wessel2012prb}. Importantly, the planar pyrochlore lattice is recently experimentally realized in the Rydberg atoms array \cite{Zoller2014prx,Lukin2021nature} and superconducting qubits systems \cite{Nisoli2013rmp,King2020}. Thus, as an experimentally achievable mixture of flat band physics and emergent low dimensional gauge field, the quantum many body physics in the planar pyrochlore lattice becomes more appealing.

%'t Hoft anormaly
On the other hand, the rapid developments of field theory provide some quantitative approaches, especially the Lieb-Schultz-Mattis-Oshikawa-Hastings (LSMOH) theorem which restricts the possible ground states \cite{LSM1961,LSM2000,LSM2004}. Recently, it is further reinterpreted as one type of mixed 't Hooft anomaly \cite{SRyu2017prb,Metlitski2018prb,CMJian2018prb,Tanizaki2018prb}. And such non-perturbative anomaly matching condition becomes useful in the contexts of deconfined quantum criticality \cite{ChongWang2017prx,LJZou2021prx,Metlitski2018prb} and also symmetry protected topological phases \cite{XGWen2013prd,XGWen2015prb,JWang2018prx}.

%short summary
In this manuscript, we study the dynamics of the half-filled extended hard-core Bose-Hubbard model in the planar pyrochlore lattice by combining quantum Monte Carlo (QMC) simulations and stochastic analytic continuation (SAC) \cite{Shao2017prb}. As shown in Fig.\ref{fig1}, away from the transition point, a flat band with bow-tie structure can be found in the low energy part and well understood with analysis of the symmetries. As an indication of fractional excitation, the continuous spectra stably appears in the high energy part. Especially close to the transition point (TP), a gapless continuous spectra can be observed along the path $(0,2\pi)\to(0,0)$ in the momentum space. These exotic phenomena can be well explained by the mixed 't Hooft anomaly mechanism, and may herald an alternative deconfined criticality.

%Model and the two limits
\emph{Model and effective theory.}---In the spin language, the model is equivalent to spin half XXZ model via conventional mapping $S_{i}^{+}\to b_{i}^{\dagger}, S_{i}^{-}\to b_{i}, S_{i}^{z}\to n_{i}-1/2$: 
\begin{equation}\label{xxzmo}
H=-t\sum_{\langle i,j\rangle}(S^{+}_{i}S_{j}^{-}+S_{i}^{-}S^{+}_{j})+V\sum_{\langle i,j\rangle}S^{z}_{i}S^{z}_{j},
\end{equation}
where $t, V>0$ are the magnitudes of XY and Ising interactions (set $V$=1 as energy unit), $\langle i,j\rangle$ represents the nearest-neighbor bond and the connected diagonal bond, as shown in Fig.\ref{fig1} (a). When $t/V=0$, the ground state is restricted by the local constraint $S_{\boxtimes}^{z}=0$ ($S_{\boxtimes}^{z}=\sum_{i\in\boxtimes}S_{i}^{z}$ is defined in a crossing plaquette ($\boxtimes$)), because the Ising interaction term can be rewritten as $\frac{V}{2}\sum_{\boxtimes}(S_{\boxtimes}^{z})^2$. Analog to the two-in-two-out ice rule in spin ice, the spins in each crossing plaquette should satisfy the two spin up and two spin down condition, so that the ground state is extensively degenerate. In the weak coupling limit $t\gg V$, the spins tend to align in the xy plane and construct the ferromagnetic phase which breaks the U(1) symmetry (superfluid (SF) phase in bosonic representation). 

In the strong coupling region $t$$\ll$$V$, the effective perturbative ring exchange interaction  $H_{\textrm{ring}}=-\frac{4t^{2}}{V}\sum_{\square}(S^{+}_{i}S^{-}_{j}S^{+}_{k}S^{-}_{l}+h.c.)$ on the non-crossing plaquette ($\square$) takes the main effect \cite{Shannon2004prb,Fradkin2006prb}, so the ground state has to maximize the number of the flippable plaquette. Then, the PVBS phase appears with spontaneous translational symmetry breaking. Indeed, the low energy physics can also be understood with help of the lattice gauge theory \cite{Hermele2004prb,Fradkin2006prb,Gingras2014review}. As demonstration of Fig.\ref{fig1}(a), the effective electric field can be defined between nearest neighbor sites in dual lattice by $E_{\textsl{rr}\,'}=\epsilon_{\textsl{r}}S_{\textsl{rr}\,'}^z$, where $\epsilon_{\textsl{r}}=+1(-1)$ in dual sublattice $\mathcal{A} (\mathcal{B})$, and the vector potential $a_{\textsl{rr}\,'}$ is  introduced by $S^{\pm}_{\textsl{rr}\,'}=\exp(\pm i\epsilon_{\textsl{r}}a_{\textsl{rr}\,'})$. Then the ice rule is equivalent to the Gauss' law: $Q_{\textsl{r}}=\textrm{div}E=\epsilon_{\textsl{r}}S_{\boxtimes}^{z}$ in which $Q_{\textsl{r}}$ is the gauge charge at dual site $\textsl{r}$. Basing on the U(1) lattice gauge theory in 2+1 dimension, the gauge charges are confined in the PVBS phase without gapless photon \cite{Polyakov1977,Fradkin2006prb}.

%'t hoft anomaly
The transition point between PVBS and SF phases is numerically found around $t/V\approxeq0.065$ \cite{Ralko2010prl}. Similar to the XXZ model in the Kagome lattice \cite{zhang_dqcp}, the system can be described with the easy-plane $\mathrm{CP}^{1}$ model or  Abelian-Higgs model \cite{Senthil2004prb,NvsenMa2018prb,Zohar2018prb}:
\begin{equation}\label{ahm}
\mathcal{L}=-\frac{1}{4e^{2}}|da|^{2}+\sum_{i=1}^{2}|D_{a}\phi_{i}|^{2}+m^{2}\sum_{i=1}^{2}|\phi_{i}|^{2}+\lambda(|\phi_1|^{2}-|\phi_{2}|^{2})^{2},
\end{equation}
where the first term is the Maxwell term \cite{Hermele2004prb,Fradkin2006prb}, $D_{a}$ is the covariant derivative and $\phi_{i}$ is the matter field. The global symmetries can be listed as following: (1) U(1) symmetry for both charges; (2) $\mathbb{Z}_{2}$ exchange symmetry $\phi_{1}\leftrightarrow \phi_{2}$; (3) $\mathcal{C}$ charge conjugation symmetry; and (4) $\mathrm{U}(1)_{\mathrm{top}}$ topological symmetry relates to the monopoles. To describe the two fold degenerate PVBS phase, the even monopoles term should be included \cite{Sandvik2017prl,Zohar2018prb}, so that the $\mathrm{U}(1)_{\mathrm{top}}$ symmetry is broken to $\mathbb{Z}_{2,\mathrm{top}}$. Then, the Lagrangian Eq.(\ref{ahm}) can hold the mixed 't Hooft anomaly among the global symmetries. When the domain wall excites between two degeneracy ground states of PVBS phase (vacuum state), it also carries anomaly due to the anomaly inflow mechanism. According to the semiclassical analysis, the theory of anomaly makes the following prediction: (1) the domain wall should break the symmetry or be gapless; (2) the domain wall (point-defect) on the domain wall (line-defect) should be deconfined; and (3) the critical theory satisfied this anomaly of the domain wall is supposed to be a compact c=1 boson conformal field theory \cite{Sandvik2017prl,Zohar2018prb}.

%Method
\emph{Phases and Bow-tie flat band}---With the improvement of the QMC-SAC technique \cite{Shao2017prb,Sandvik2016pre}, the dynamical structure factor (DSF) $S^{zz}(\bm{q},\omega)$ plays a same distinctive role in the numerical simulation as the neutron inelastic scattering in the experiment \cite{NIS_2012}, which can detect the exotic quantum phase, but also the excitations especially with the topological features \cite{Zhao2019np,BFG_Meng,BFG_Wessel}. Hence, we adopt the QMC-SAC method to simulate the dynamics of the system which hosts the emergent U(1) lattice gauge field and fractional excitations. In order to achieve the high resolution spectrum reflecting low energy physics, the lattice size is set to $N=32\times32=1024$ sites with periodic boundary condition and the temperature is lower to $T=1/\beta\approx t^{2}/V$.
\begin{figure}
	\centering
	\includegraphics[width=0.48\textwidth]{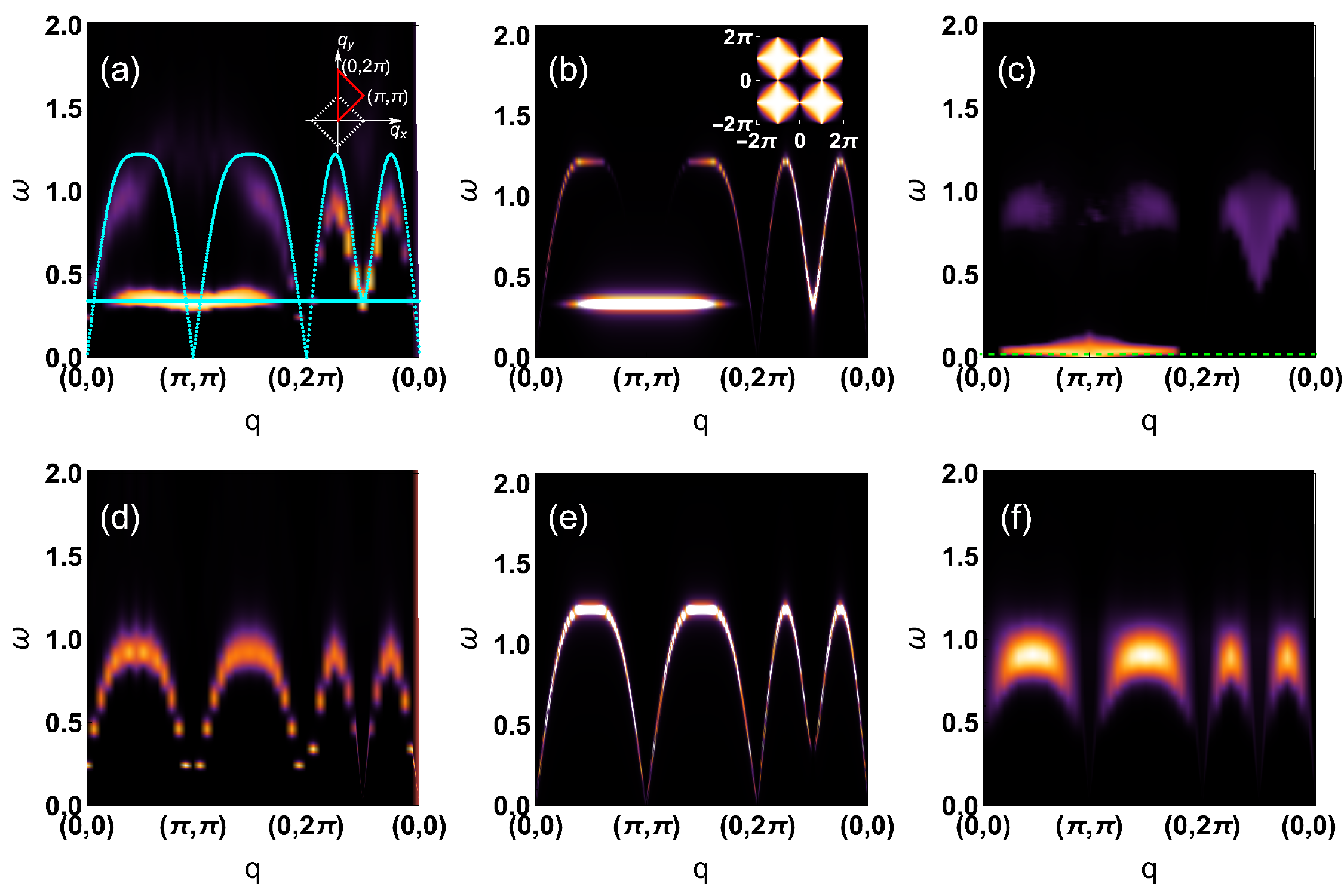}
	\caption{\label{fig2} The DSF $S^{zz}(\bm{q},\omega)$ (a-c) and the charge-charge correlation spectra $D(\bm{q},\omega)$ (d-f) in SF phase at $t/V$=0.1 (a,b,d,e) and PVBS phase at $t/V$=0.05 (c,f). The spectra of (b,e) are calculated from SWT, while others are from QMC-SAC. The cyan sold lines in (a) are energy dispersion calculated by SWT. The inset in (b) is the constant energy cut of the DSF calculated by SWT with energy$\sim0.34V$ very close to the flat band. The green dashed line in (c) at $\omega\approxeq0.02$ is calculated by POT.}
\end{figure}

%The DSF in SF phase
The DSF in SF phase with $t/V$=0.1 is shown in Fig.\ref{fig2}(a), and it consists of two branches. The higher energy branch exhibits strong momentum dependent dispersion, and a flat band can be clearly observed at the lower energy. However, different from spreading to the whole Brillouin zone, the flat band vanishes along the $\textbf{q}$: $(0,2\pi)\rightarrow(0,0)$. To have a better understanding, we make use of the linear spin wave theory (SWT), and take $t/V$=0.18 to fit the numerical spectrum, because SWT usually underestimates the quantum fluctuation especially in the geometrical frustrated system \cite{Auerbach1997prb}. The dispersion can be calculated by diagonalizing the Hamiltonian of the quadratic Holstein-Primakoff bosons $a_{j,\alpha}$, where $\alpha$ is the A,B sublattice index. In the  Fig.\ref{fig2}(a), both flat band and dispersive branches got by SWT (cyan lines) match well with the numerical results. Meanwhile, the DSF from SWT shown in Fig.\ref{fig2}(b) can reproduce many features of the numerical spectra, especially the fragmentation of the flat band. Importantly, after focusing on the energy where flat band stays, a \textit{bow-tie} structure is revealed in the inset of Fig.\ref{fig2}(b), and it hints possible ``selection rule" related to the symmetry.

%Analysis of the group theory
The planar pyrochlore lattice has the glide reflection symmetries $G_{x}$:$(x,y)\to(x+1,-y)$ and $G_{y}$:$(x,y)\to(-x,y+1)$ \cite{JYLee2019prx}. Correspondingly, the Holstein-Primakoff bosons follows the transformation: $G_{x}a_{(q_x,q_y),\alpha}G_{x}^{\dagger}=e^{iq_{x}}a_{(q_x,-q_y),\bar{\alpha}}$, $G_{y}a_{(q_x,q_y),\alpha}G_{y}^{\dagger}=e^{iq_{y}}a_{(-q_x,q_y),\bar{\alpha}}$, and $\bar{A}=B$, $\bar{B}=A$. Then, the two eigenmodes of $G_{y}$ along the $q_{x}=0$ path take the form \cite{CLLiu2018prb} $g_{(0,q_y),\pm}=a_{(0,q_y),A}\pm a_{(0,q_y),B}$, so that the eigenmodes of the Hamiltonian taking Bogoliubov form changes to $b_{(0,q_y),\pm}=u_{(0,q_{y}),\pm}g_{(0,q_y),\pm}+v_{(0,q_{y}),\pm}g_{(0,-q_y),\pm}^{\dagger}$. In real space, the minus mode has alternating sign and constant modulus for sublattice A and B around the non-crossing plaquette, this is known to be the flat mode and origins from the localized excitation \cite{Mielke2015,Harris1992prb}. Considering $S^{z}_{\bm{q}}\propto a_{\bm{q},A}+a_{-\bm{q},A}^{\dagger}+a_{\bm{q},B}+a_{-\bm{q},B}^{\dagger}$ in the SWT, the DSF can only reflect the plus mode. Therefore, the magnitude of the DSF $S^{zz}(\bm{q},\omega)$ on the flat band totally vanishes along $q_{x}=0$ and $q_{y}=0$.

%Flat band in VBS
Same phenomena can also be observed in the PVBS phase, as shown in Fig.\ref{fig2} (c) with $t/V$=0.05, the flat band branch stays along $\textbf{q}$: $(0,0)\rightarrow(\pi,\pi)\rightarrow(0,2\pi)$ at very low energy and is separated from the high energy continuum by a large gap. Due to the order from disorder mechanism, the ring exchange interaction $H_{\textrm{ring}}$ can lift the degeneracy and break the translational symmetry. Thus, the energy scale of the flat band is about $4t^{2}/V$. A better estimation can be made by plaquette operator theory (POT) \cite{Langari2015epjb}. In the leading order, we show the dispersion of the lowest excitation with the green dash line in Fig.\ref{fig2} (c). It is flat with energy $\omega\approxeq0.02$ and corresponds to the excitation from the ground state of an isolated non-crossing plaquette $|e_{g}\rangle\approx|\downarrow\uparrow\downarrow\uparrow\rangle+|\uparrow\downarrow\uparrow\downarrow\rangle$ (valid when $t\ll V$) to the first excited state $|e_{1}\rangle=|\downarrow\uparrow\downarrow\uparrow\rangle-|\uparrow\downarrow\uparrow\downarrow\rangle$. Considering the flat band also vanishes along $k_{x}=0$ in PVBS, we think such ``selection rule" or fragmentation of flat band may be universal and due to the incompatibility between DSF and flat band physics.

The high energy branches of the PVBS are at the energy scale of $V$ which indicates it may be related to the violation of the ice rule \cite{zhou2020}. The charge density can be defined as $m_{\textsl{r}}=\sum_{i\in\boxtimes_{\textsl{r}}}S_{i}^{z}$, and the ice rule can restrict the Hilbert space to the pure gauge field sector without charge excitation. Then, the quantum fluctuation of emergent gauge field contributes to the flat band. However, the spin exchange interaction can break the ice rule and produce two gauge charges in the same sublattice costing energy $V$. In order to study the contribution of the gauge charges on the spectrum, we numerically simulate the spectrum of charge density-density correlation $D(\bm{q},\omega)=\int^{\infty}_{-\infty}dt \langle m_{\bm{q}}(t)m_{-\bm{q}}(0)\rangle e^{i\omega\,t}$, and also calculate it in SF phase with SWT. From Fig.\ref{fig2} (d-f), we can find the disappearance of the flat band, and it means the high energy branches directly result from the excitation of gauge charge. In the SF phase, the comparison of results from SWT and QMC demonstrates the dispersion of the charge-charge correlation is the same as the dispersive spin wave band. On the other hand, both DSF and gauge charge spectrum in the PVBS phase exhibit similar dispersive features, but also a distinct continuum along the $k_x=0$ which is usually related to the fractionalization \cite{NIS_2012}.

%The phase transitions 
\emph{Phase transition and 't Hooft anomaly.}---The PVBS and SF phases break different symmetries, so the phase transition between them should be first order based on the Ginzburg-Landau theory. However, it also could be continuous due to the deconfinement of the topological defects \cite{Senthil2004science}. Meanwhile, even if it is first order, the recent work on the checkerboard J-Q model demonstrates the effective gauge field still plays a critical role \cite{Zhao2019np}. Thus, rather than interminable finite size scaling of the ground state, analyzing the features of the spectrum around the phase transition point may provide more clues to the physical mechanism.

\begin{figure}
	\centering
	\includegraphics[width=0.48\textwidth]{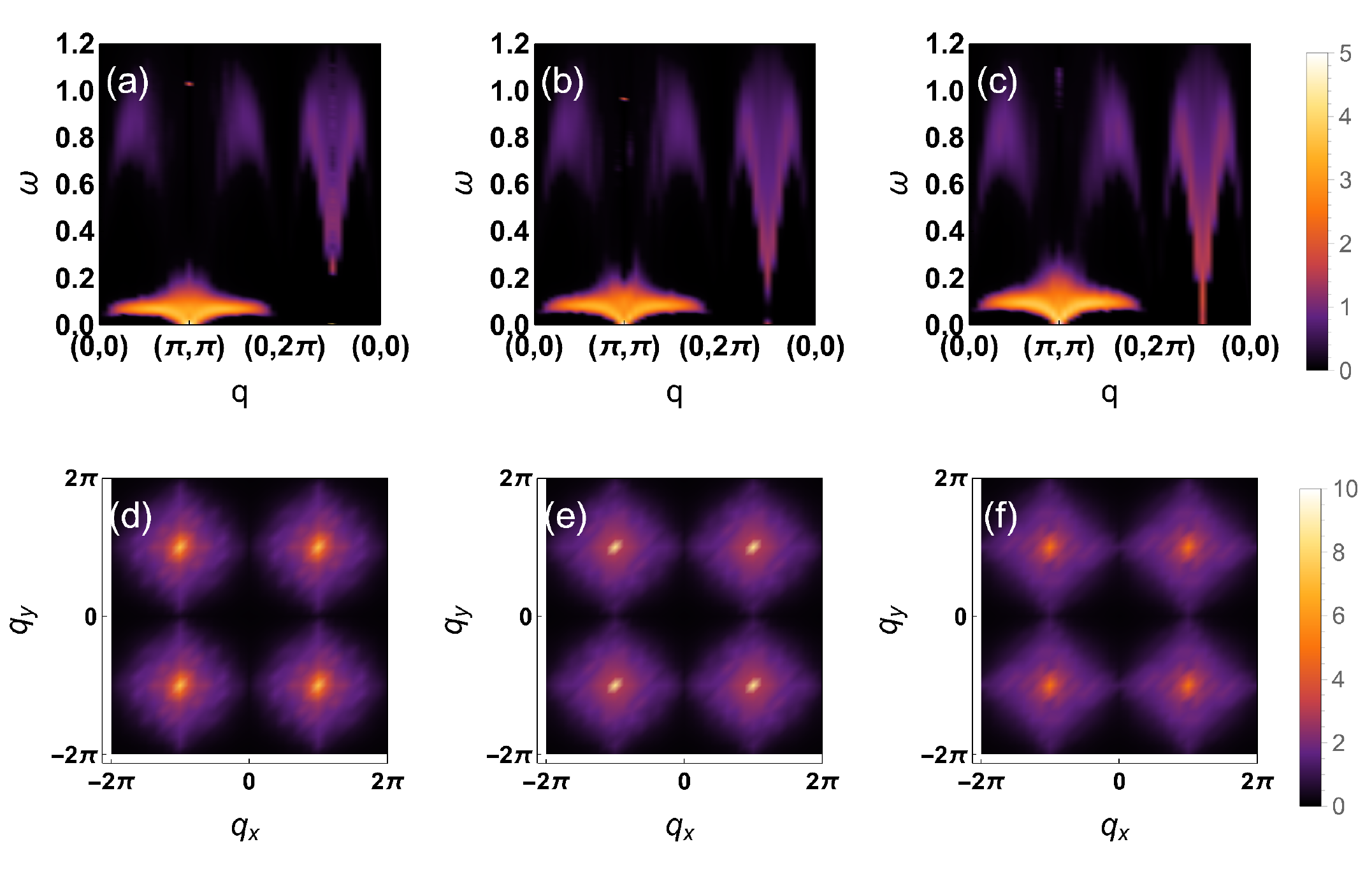}
	\caption{\label{fig3} DSF $S^{zz}(\bm{q},\omega)$ (a-c) and static structure factor $S^{zz}(\bm{q})$ (d-f) close to the transition point at $t/V$ equal to 0.06, 0.063, and 0.065 from left to right.}
\end{figure}
%The continous phase transition near the critical point
While approaching the transition point from VBS side, the flat band turns to be gapless at $(\pi,\pi)$ in Fig.\ref{fig3} (a-c). Remarkably, the high energy branch at $(0,\pi)$ becomes continuous along with falling down to zero at transition point. These exotic features are reminiscent of the gapless emergent fractional excitations and the deconfined quantum phase transition \cite{Senthil2004science,NvsenMa2018prb,zhang_dqcp}. And further from the static structure factor $S^{zz}(\bm{q})=\langle |S^z(q)|^2\rangle$ in Fig.\ref{fig3} (d-f), as the hallmark of translational symmetry breaking, the high intensity at four high symmetry points  $(\pm\pi,\pm\pi)$ in PVBS phases changes to be blur at transition point, and the bow-tie structure indicates the power law correlation with gapless excitations \cite{Henley2010,Gingras2014review}. However, the conventional theory of the deconfined quantum criticality \cite{Nahum2015prl} can not support an emergent symmetry $\rm{U}(1)\times \mathbb{Z}_{2,\mathrm{top}}\to\rm{O}(3)$ in which $\rm{U}(1)$ and  $\mathbb{Z}_{2,\mathrm{top}}$ are from SF and PVBS phases, respectively. Then, the deconfined domain wall (zero dimension topological defect) on the domain wall (one dimension topological defect) from the 't Hooft anomaly looks like an alternative possible scenario \cite{Sandvik2017prl}.

\begin{figure}[t]
	\centering
	\includegraphics[scale=0.5]{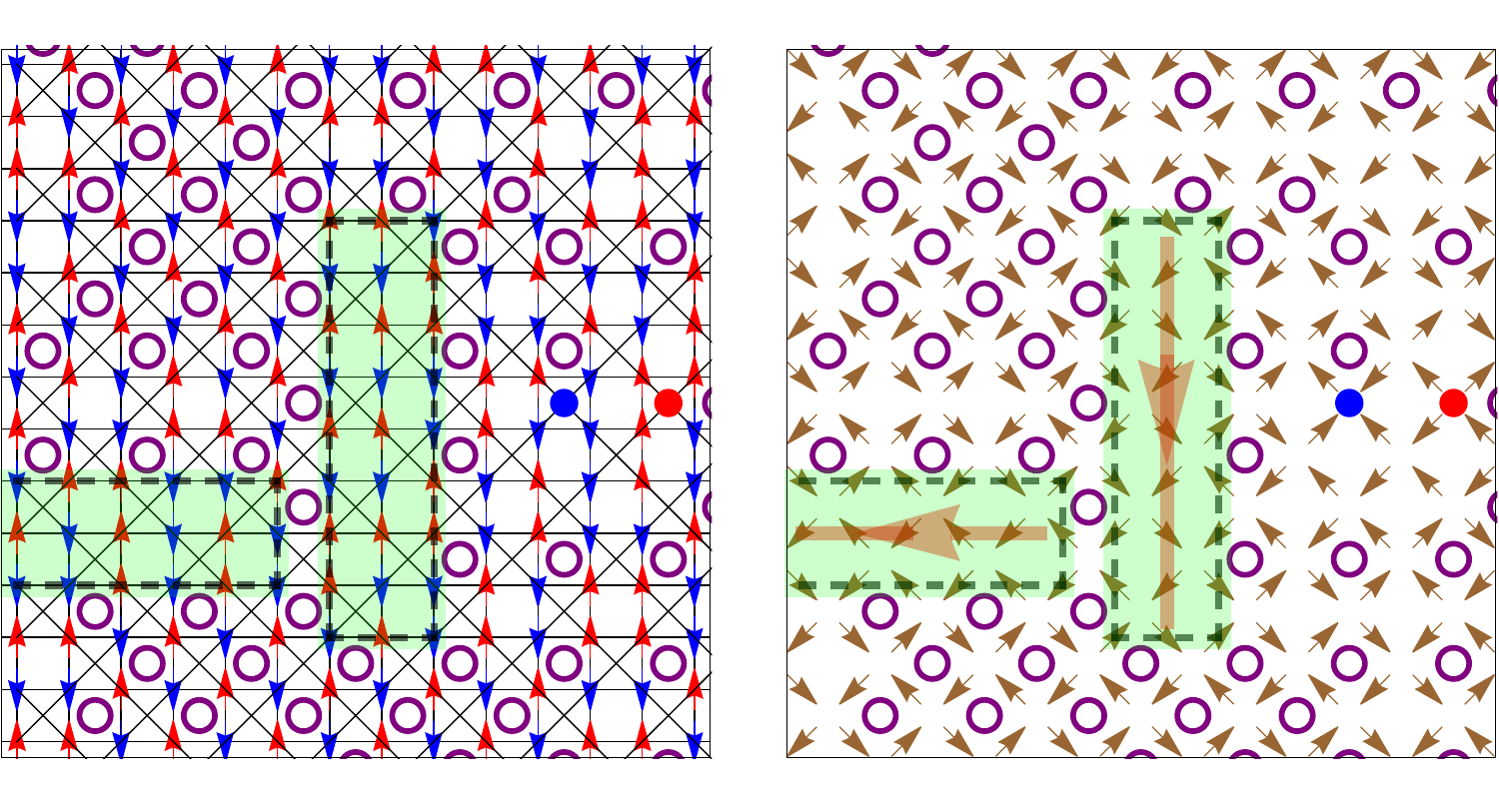}
	\caption{\label{fig4}(Left) The snapshot of the spin configuration from the QMC simulation at $t/V=0.05$. Spin up and down are labeled with red and blue arrows, respectively. The purple circles mark the flippable non-crossing plaquette. The domain walls are highlighted with a green region. (Right) The corresponding effective electric field representation of the left snapshot. The big red arrow denotes the direction of the electric field in the domain wall.}
\end{figure}
%snapshot
The effective field theory presents the two fold degeneracy of PVBS phase corresponding to the symmetry $\mathbb{Z}_{2,\mathrm{top}}$, and they refer to one sublattice of non-crossing plaquette is flippable. When the domain wall excites between two vacua, it can restore the $\mathbb{Z}_{2,\mathrm{top}}$ and break the $\mathcal{C}$ charge conjugation symmetry. Similar to the near-field measurement in the experiment, we make a snapshot of the spin configuration in the PVBS phase which can easily demonstrate the structure in the real space \cite{zhang_ss}. As shown in the left panel of Fig. \ref{fig4}, two different kinds of domain walls (green region) can be formed between two vacua, and they can only lie in the x or y direction. Because all the non-crossing plaquettes in the domain walls are not flippable, the symmetry $\mathbb{Z}_{2,\mathrm{top}}$ is restored. Then turning to the effective electric field picture in the right panel of Fig. \ref{fig4}, the electric fields in the domain walls are polarized (marked with big red arrow). It indicates the domain wall breaks the charge conjugation symmetry because they carry electric flux, and these features are consistent with the semiclassical analysis of 't Hooft anomaly \cite{Zohar2018prb}.

\begin{figure}[h]
	\centering
	\includegraphics[scale=0.48]{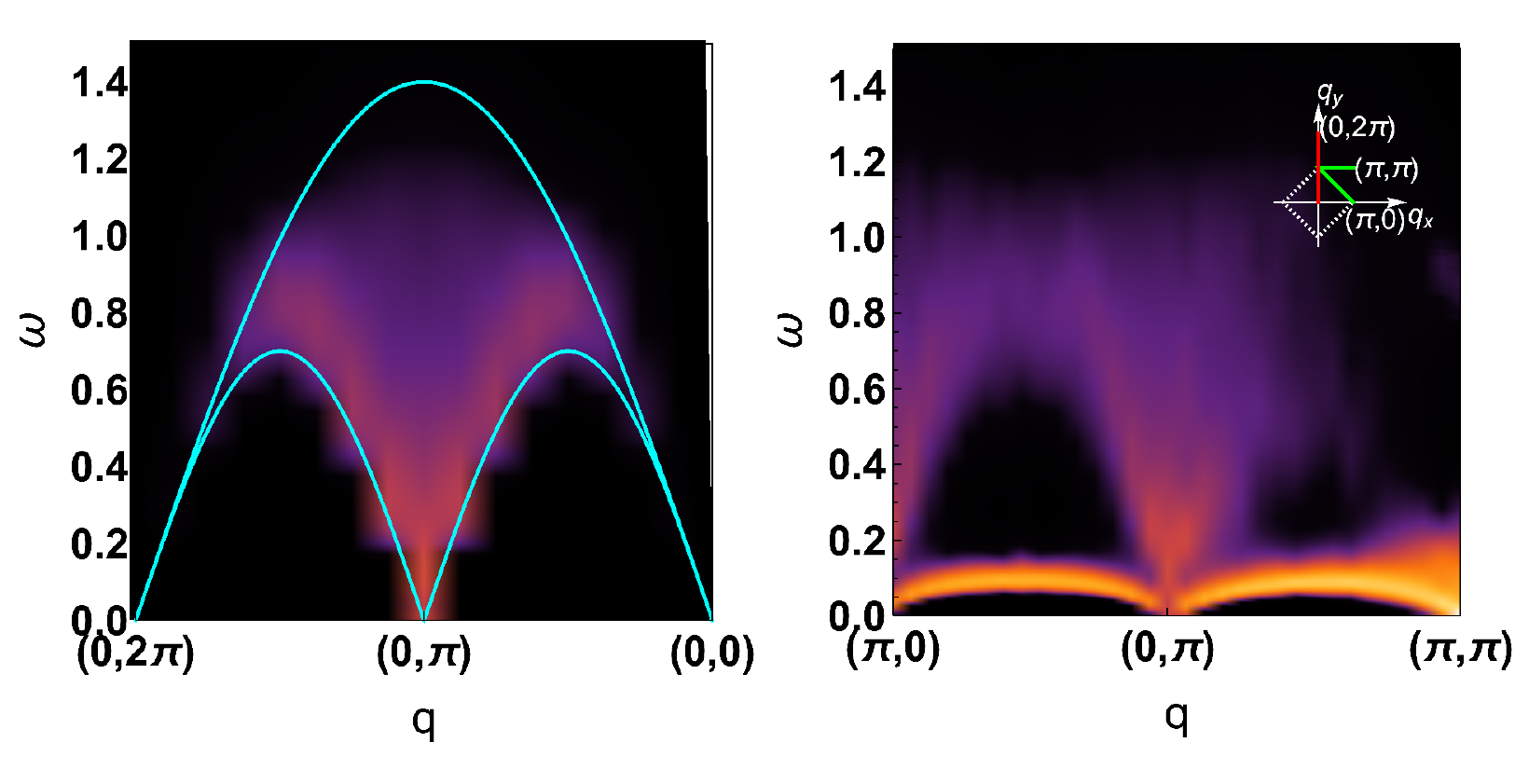}
	\caption{\label{fig5} DSF at the transition point along different momentum paths. The cyan lines are fitted by the upper and lower boundaries of two spinon continuum of XXZ spin chain. Inset: the red and green lines in the momentum space label the paths of left and right panel, respectively.}
\end{figure}
%deconfined domain wall
At the transition point, the snapshot method fails because the domain walls are intertwined with each other. However, the specific directions of domain walls determine that the corresponding part in the spectra should be anisotropic. Meanwhile, the domain wall excitations on the domain wall should be gapless and exhibit the characteristics of conformal field theory \cite{Zohar2018prb}. Thus, we zoom in the spectra of Fig.\ref{fig3}(c) along $(0,2\pi)\to (0,0)$ in left panel of Fig.\ref{fig5}, and the gapless continuum stays between the two envelope curves (cyan lines) which are the lower and upper boundaries of the two spinon continuum of the critical XXZ spin chain \cite{Caux2011prl}. The critical theory of XXZ spin chain is compact boson c=1 conformal field theory, and it matches well with predication of mixed 't Hooft anomaly. Furthermore, another path along $(\pi,0)\rightarrow(0,\pi)\rightarrow(\pi,\pi)$ in the right panel of Fig.\ref{fig5} demonstrates the continuum is strongly anisotropic, and the closing of the energy gap involves the interplay between the emergent gauge field (flat band and domain walls). Then at the transition point, although the deconfinement is forbidden in bulk, it can still be allowed on the domain wall. It is noteworthy that this "confinement in the bulk, deconfinement on the wall" phenomenon is also found in the $\mathbb{Z}_{2}$ VBS phase in the $J-Q_{x}-Q_{y}$ model on the square lattice recently \cite{Sandvik2017prl,Zohar2018prb}. In contrast to that, this mechanism serves as an alternative deconfined transition mechanism in our model, and it is natural to expect the enhanced symmetry also exists at this SF-PVBS transition point.

\emph{Conclusions.}---We studied the spectra of the half-filled extended hard-core Bose-Hubbard model (equivalent to the spin half XXZ model without external field) in the planar pyrochlore lattice. In both PVBS and SF phases, we observe the fragmentation of the flat band. With the analysis of the symmetries, we found the bow-tie flat band is due to the local resonant process and the lattice geometry. It means the fragmentation of the flat band may be intrinsic and exist in other flat band geometry \cite{Xiong2020npj}, e.g. Kagome lattice. The energy gap is closed at the transition point, and an emergent critical continuum is found at $(0,\pi)$. The physical mechanism can be understood with the effective Abelian-Higgs model. As the prediction of previous semiclassical analysis \cite{Zohar2018prb}, many characteristics of the mixed 't Hooft anomaly are founded. The domain wall exists between two vacua and restores the $\mathbb{Z}_{2,\mathrm{top}}$ symmetry, but break the charge conjugation symmetry because it must maintain the anomaly. The domain wall appears only in x or y direction, so the spectra is strongly anisotropic around $(\pi,\pi)$. Last but not least, we found the Luttinger liquid like continuum in the spectra along the  $k_{x}=0$, which hints the existence of gapless domain wall excitations on the domain wall. Thus, we think such deconfinement mechanism may be an alternative interpretation of the continuous phase transition beyond the Ginzburg-Landau paradigm.
%%%%%%%%%%%%%%%%%%%%%%%%%%%%%%%%%%%%%%
We thank Yuan Wan, Fa Wang, Long Zhang, Yi-Zhuang You and Changle Liu for helpful discussions. We acknowledges funding from the National Science Foundation of China under Grants No. 11804034, No. 11874094 and No.12047564, Fundamental Research Funds for the Central Universities Grant No. 2020CDJQY-Z003 and 2021CDJZYJH-003.

%%%%%%%%%%%%%%%%%%%%%%%%%%%%%%%%%%%%%
\bibliographystyle{apsrev4-1}
\bibliography{ref}
\end{document}